\documentstyle[twocolumn, prb,aps]{revtex}
\begin{document}
\draft
\begin{title}
{Angle-resolved photoemission in doped charge-transfer Mott insulators}
\end{title} 
\author{A.S. Alexandrov and C. J. Dent}
\address
{Department of Physics, Loughborough University, Loughborough LE11 
3TU, U.K.}

\maketitle
\begin{abstract}
A theory of angle-resolved photoemission (ARPES) 
in doped cuprates and 
other charge-transfer Mott insulators is developed taking 
into account the realistic (LDA+U) band structure, (bi)polaron formation 
due to the
strong electron-phonon interaction, and a random field potential. 
In most of these materials 
the
first band to be doped is the oxygen band inside the Mott-Hubbard gap. We 
derive the coherent part of the ARPES spectra with the oxygen hole spectral function 
calculated in  
the non-crossing (ladder) approximation   
and with the exact spectral function of a one-dimensional 
hole in a random potential. Some unusual features of ARPES including 
the polarisation dependence and spectral shape in  
YBa$_{2}$Cu$_{3}$O$_{7}$ and 
YBa$_{2}$Cu$_{4}$O$_{8}$ 
are described without 
any Fermi-surface, large or small. The theory 
 is compatible with the doping dependence of kinetic and 
thermodynamic properties of cuprates as well as with the d-wave symmetry 
of the superconducting order parameter.

\end{abstract}
\narrowtext

\centerline {\bf 1. Introduction}

 The concept of polarons led to the discovery of 
  the copper oxide superconductors. 
The expectation was that if `an electron and a surrounding lattice distortion 
with a high effective mass can travel through the lattice as a whole, and 
a strong electron-phonon coupling exists, the perovskite insulator could 
be turned into a high temperature superconductor' \cite{bed}. In this 
paper we develop the theory of ARPES in doped charge-transfer Mott 
insulators based on the bipolaron theory 
\cite{alemot,aleF} which  describes some unusual ARPES features of  
 high-T$_{c}$  YBa$_{2}$Cu$_{3}$O$_{7-\delta}$ (Y123),  
 YBa$_{2}$Cu$_{4}$O$_{8}$ (Y124) and a few other materials. 
 
 ARPES data in cuprates remains highly controversial \cite{rev}. 
 One of the  surprising features is a  $large$ Fermi surface
 claimed to exist in a wide range of doping  fitting well the 
 LDA band structures in the earlier studies. This should evolve with doping as $(1-x)$ 
 in a clear 
contradiction with low frequency kinetics and thermodynamics (see, for 
example \cite{jon,mul2,mic,hof,bat,lor}, which show 
an evolution proportional to $x$ in a wide range of doping including 
the overdoped region \cite{bat,opt} ($x$ is the number of holes introduced by doping). Now it is 
established, however, that there is a normal state (pseudo)gap in ARPES and  
tunnelling, existing well above 
T$_{c}$ \cite{sch,ren}, so that some  segments of a `large Fermi surface' 
are actually missing \cite{bia}. The  temperature and doping 
dependence of the gap still remain a subject of controversy. While  kinetic 
\cite{bat}, thermodynamic \cite{lor}, tunnelling \cite{ren} and some 
ARPES \cite{bia,bia2} data suggest that the gap opens at any relevant
temperature in a wide range of doping, other ARPES studies 
\cite{din,rand,cam}
claim that it exists only in underdoped samples  below a 
characteristic temperature T$^*$. 
 
Perhaps the most intriguing feature of ARPES is the extremely narrow and 
intense peak lying below the Fermi energy, which is most clearly 
seen near the Y and X 
points in Y124\cite{gof}, Y123 \cite{mat} and more recently in 
La$_{2-x}$Sr$_{x}$CuO$_{4}$ \cite{ino}. Its angular dependence and
 spectral shape as well as the origin of the featureless (but dispersive) 
background remain unclear. Some authors \cite{gof} refer to the peak
as an extended van Hove singularity (evHs) arising from the plane 
(CuO$_2$)
strongly correlated band. They also implicate the resulting 
(quasi)1D density of states singularity as a possible origin for the high 
transition temperature. However, recent polarised ARPES studies of untwinned Y123 
crystals of exceptional quality \cite{mat} unambiguously refer the peak to 
as a narrow resonance arising primarily from the quasi-1D CuO$_3$ chains 
in the buffer layers rather than from the planes. Also, careful analysis  
of the Eliashberg equations shows that 
the van Hove singularity can hardly  be the origin of  high T$_c$, 
 in sharp contrast with a 
naive weak-coupling estimate. Interestingly, a very similar narrow peak 
was observed by Park $et$ $al$ \cite{par} in high-resolution ARPES
 near the gap edge of the cubic
$semiconductor$ FeSi with no Fermi surface at all.

In this paper we take the view that cuprates and many other transition 
metal compounds are charge transfer Mott-Hubbard insulators at $any$ level 
of doping \cite{aleR}. This means that  the first band to be doped is the 
oxygen band lying within the Hubbard gap, Fig.1.  
Strong coupling with  high frequency phonons, unambiguously established 
for many oxides \cite{mul}, leads to 
 the  high-energy spectral features of an oxygen hole
in an energy window about twice the Franck-Condon (polaronic) level shift,
 $E_p$, and to 
the band-narrowing effect \cite{alemot}. On the other hand,
the low energy spectral function is influenced
  by the low frequency thermal lattice, spin and random  fluctuations. 
The latter can be described as `Gaussian white noise'. 
The $p$-hole polaron in oxides is almost 
one-dimensional due to a large difference in the $pp\sigma$ and $pp\pi$ 
hopping integrals and effective `one-dimensional'
localisitaion  by the random 
potential as described in Ref.\cite{ale2}. 
  This allows us  to explain  the narrow peak in the ARPES spectra with 
the  spectral density $A(k,E)$ of a one-dimensional particle in a
Gaussian white noise potential\cite{hal}. 

The 
electron-phonon interaction also binds  holes into inter-site oxygen 
 bipolarons the size of a lattice constant \cite{alemot,cat2}. 
The bipolaron density remains  relatively 
low (below 0.15 per cell) at any relevant level of doping. The residual 
repulsive (Coulomb) polaron-bipolaron interaction \cite{alemot}
 is strongly suppressed 
by the lattice  polarisation  owing to a large 
 static dielectric constant. As a result, the only role of hole 
 bipolarons in ARPES is 
to pin the chemical potential inside the charge 
transfer gap, half the bipolaron binding 
energy above the oxygen band edge, Fig.1a. This binding energy as well 
as the singlet-triplet bipolaron exchange energy are thought to be 
the origin of the normal state pseudogaps, as first proposed by one of 
us \cite{ale3}. 
In particular, the non-Korringa temperature dependence of 
NMR \cite{ale3}, and more recently the values 
and universal scaling with temperature of the uniform magnetic 
susceptibility \cite{alekabmot,mul2,alekay} and electronic specific heat 
\cite{alekay} were quantitatively described with bipolarons.
    In overdoped samples the bipolaron and polaron bands might overlap 
because the bipolaron binding energy becomes small \cite{alekabmot}, so the chemical 
potential might enter  the oxygen band, Fig.1b, and then a Fermi-level 
crossing might be seen in  ARPES.
The 
featureless background is explained as  the phonon cloud of a small 
hole polaron, which spreads over a wide energy interval about $2E_p 
\simeq 1eV$. 

The same approach has  been  applied  to the   tunnelling spectra 
\cite{aletun}.
As a result, the temperature independent gap and the 
anomalous $gap/T_{c}$ ratio, injection/emission asymmetry both in  
magnitude and  shape, zero-bias conductance at zero temperature, the spectral shape inside and 
outside the gap region, temperature/doping dependence 
and dip-hump structure  of the tunnelling conductance were described.
 The approach is compatible with the recent `LDA+U' band structures in 
 cuprates and manganites \cite{saw}, which suggest the 
 single-particle density of 
 states shown in Fig.1.  It is  clearly 
compatible with the doping evolution of thermodynamic and kinetic 
properties because holes introduced by doping into the oxygen band are
the only carriers in our theory.  
Moreover, the derived $bipolaron$ energy dispersion with minima at the 
Brillouin zone boundaries provides  $d$-wave symmetry of the 
Bose-Einstein condensate  \cite{aletun} in agreement with  
 phase-sensitive experiments \cite{pha}. It also provides
 a parameter-free 
expression  for T$_c$ in a few dozen cuprates, irrespective of the 
 level of doping \cite{alecom,alecom2}. 
 
 Here we  first derive  the polaronic 
 ARPES theory in Section 2. Then, in Section 3 we apply the simplest non-crossing approximation 
 for the  self-energy and compare it with 
 the exact result in Section 4. The experimental ARPES results, in particular 
 the 
 angular dispersion, spectral shape and polarisation dependence 
 are compared with the theory in Section 5. We conclude  that the narrow 
 peak in ARPES  is 
 an $intrinsic$ polaron (oxygen) band found below the chemical 
 potential by half of the bipolaron binding energy. We also arrive at 
 the conclusion that the present experimental photoemission spectra can be understood 
 with a small  or even $without$ $any$ Fermi surface (depending on 
 doping).

\centerline {\bf 2. Polaronic ARPES}

The interaction of the crystal with the electromagnetic field of 
frequency $\nu$   is described by the 
Hamiltonian (in the dipole approximation)
\begin{equation}
H_{int}= (8\pi I)^{1/2} sin(\nu t) \sum_{\bf k,k'} ({\bf e }
\cdot {\bf d}_{\bf k,k'}) c^{\dagger}_{\bf k}h^{\dagger}_{\bf k'} +H.c.,
\end{equation}
where $I$ is the intensity of the radiation with the polarisation $\bf 
e$ (we take 
$c=\hbar=k_{B}=1$), $\bf k$ is the momentum of the final state (i.e. of 
the photoelectron 
registered by the 
detector), $\bf k'$ is the (quasi)momentum of the 
hole remaining in the sample after the emission, and $c^{\dagger}_{\bf 
k}$ and $h^{\dagger}_{\bf k'}$ are their creation operators, 
respectively. For simplicity we suppress the band index in 
$h^{\dagger}_{\bf k'}$. Due to the translational symmetry of the Bloch 
states, $|{\bf k'}>\equiv u_{-\bf k'}({\bf r}) exp(-i\bf k'\cdot r)$, there is 
a momentum conservation rule in the dipole matrix element,
\begin{equation}
{\bf d}_{\bf k,k'}={\bf d}(\bf k) \delta_{\bf k+k',G}
\end{equation}
with 
\begin{equation}
{\bf d}({\bf k}) =ie (N/v_{0})^{1/2} \nabla_{\bf k} \int_{v_{0}} 
e^{{-i\bf G\cdot r}} u_{\bf 
k-G}({\bf r})d{\bf r},
\end{equation}
 where ${\bf G}$ is a reciprocal lattice vector,  $N$ the number of unit 
 cells in the crystal of the volume $Nv_{0}$. 

The Fermi Golden Rule gives the photocurrent to be
\begin{eqnarray}
I({\bf k}, E)&=&4 \pi^{2}е I |{\bf e \cdot d}({\bf k})|^{2} \cr\times\sum_{i,f}
e^{\Omega+\mu N_{i}-E_{i}}|<f&|&h^{\dagger}_{\bf 
k-G}|i>|^{2} \delta (E + E_{f}-E_{i}),
\end{eqnarray}
where $E$ is the binding energy, $E_{i,f}$ is the energy of the initial and final states of 
the crystal,  and $\Omega, \mu, N_{i}$ are the thermodynamic 
and chemical potentials and the number of holes, respectively. By definition 
 the sum in Eq.(4) is 
 $n(E)A({\bf k-G}, -E)$ where the 
spectral function $A({\bf k-G}, E)=(-1/\pi) \Im G^{R}({\bf k-G},E)$ 
is proportional to the imaginary part of the  retarded Green function (GF), and 
$n(E)=[exp(E/T)+1]^{-1}$, the Fermi distribution. In the following we consider 
temperatures well below the experimental energy resolution, so that 
$n(E)=1$ if $E$ is negative and zero otherwise, and, for 
convenience, we put ${\bf G}=0$.

 The spectral 
function depends on essential interactions of a single hole with 
the rest of the system. As we  argued earlier \cite{ale2} the most important 
interaction in oxides is the Fr\"ohlich long-range electron-phonon interaction 
of the oxygen hole with the 
c-axis  polarised high-frequency phonons.  The 
Fr\"ohlich interaction is  
integrated out with the Lang-Firsov 
canonical transformation \cite{lan}. With the 
 Quantum-Monte-Carlo technique \cite{alekor} one can prove that 
 this transformation  is
practically exact for a long-range interaction in a wide region of 
the coupling strength, including the intermediate and weak-coupling 
regime and in a wide  range of the adiabatic ratio, $\omega /D$ 
($\omega$ is the phonon frequency and $D$ is the  bandwidth).  
By applying the Lang-Firsov canonical 
transformation the hole Matsubara GF is expressed as a convolution of the 
coherent polaron GF and the multiphonon correlation function 
$\sigma({\bf m},\Omega_{n})$
\cite{aleF},
 \begin{equation}
{\cal G} ({\bf k},\omega_{n})={T\over{N}}\sum_{\omega_{n'},{\bf 
m,k'}}{\sigma({\bf m},\omega_{n'}-\omega_{n})e^{i({\bf 
k-k'})\cdot{\bf m}}\over{i\omega_{n'}-\xi_{\bf 
k'}}},
\end{equation}
where   
 the multi-phonon correlation function, $\sigma({\bf 
m},\tau)=T\sum_{n}e^{-i\Omega_{n}\tau}\sigma({\bf m},\Omega_{n})$ is found as
\begin{equation}
\sigma({\bf m},\tau) = exp \left ({1\over{2N}}\sum_{\bf q}|\gamma({\bf 
q})|^{2}
f_{\bf q}({\bf m},\tau) \right).
\end{equation}
Here
\begin{eqnarray}
\nonumber
f_{\bf q}({\bf m},\tau)&=&[cos({\bf q} \cdot {\bf m}) cosh (\omega_{\bf q}|\tau|) -1] 
coth{\omega_{\bf q}\over{2T}}
\cr&+& cos({\bf q}\cdot {\bf m}) sinh(\omega_{\bf 
q}|\tau|) 
\end{eqnarray}
with $\bf m$ the lattice vector, $\omega_{n}= \pi T (2n+1)$, $n=0,\pm 
1,\pm 2 ...$ and $\Omega_{n}=2\pi T n$.
 
 In the  case of  dispersionless phonons  and a 
 short-range (Holstein) interaction with a $q$ -independent matrix 
 element ($\omega_{\bf q}=\omega$, $|\gamma({\bf q})|^{2}=2g^{2}$)
  one can readily calculate the Fourier component of $\sigma({\bf m},\tau)$ 
   to obtain \cite{aleF}  
\begin{eqnarray}
{\cal G}({\bf k},\omega_{n}) &=& {e^{-g^{2}}\over{i\omega_{n}-\xi_{\bf 
k}}}+
{e^{-g^{2}}\over{N}}\sum_{l=1}^{\infty}{g^{2l}\over{l!}}\cr
&\times& \sum_{\bf k'}
\left({n_{\bf k'}\over{i\omega_{n}-\xi_{\bf k'}+l\omega}}+
{1-n_{\bf k'}\over{i\omega_{n}-\xi_{\bf 
k'}-l\omega}}\right).
\end{eqnarray}
The  Green's function of a polaronic carrier comprises
 two different contributions. The first coherent $\bf k$-dependent
 term arises from the 
polaron band tunneling.  The spectral weight of the coherent part is 
strongly (exponentially) suppressed as $Z=exp(-g^2)$ and the effective 
mass is strongly enhanced, $\xi_{\bf k}=Z E_{\bf k}-\mu$ (we include the 
polaronic level shift into the chemical potential, $\mu$) . Here $E({\bf 
k})$ 
is the bare (LDA+U) $hole$ band dispersion. 
The second ${\bf k}$-independent contribution 
describes the excitations accompanied by the emission and absorption of  
phonons. 
 We believe that this term, $I_{incoh}(E)$  is responsible for the 
asymmetric background in the optical conductivity and in the 
photoemission spectra of cuprates and manganites. We notice that its 
spectral density  spreads over a wide energy range 
of about  twice the polaron level shift $E_{p}=g^{2}\omega$. On the 
contrary the coherent term shows an angular dependence in the energy 
range of the
order of the polaron bandwidth $2w\equiv Z D$.

As we have discussed above ARPES measures  
the imaginary part of the retarded GF 
 multiplied by the Fermi-Dirac distribution function and by the square of 
 the dipole matrix element.
 $G^{R}$ is obtained 
 from ${\cal G}({\bf k},\omega_{n})$ with a  substitution 
  $i\omega_{n}\rightarrow E 
  +i0^{+}$. As a result we obtain
 \begin{equation}
 I({\bf k},E) \sim |d({\bf k})|^{2}n(E) Z\delta (E+\xi_{\bf 
 k})+I_{incoh}(E),
 \end{equation}
where $I_{incoh}(E)$ is a structureless  function of the binding 
energy, which spreads from about $-\omega$ down to $-2E_p$. Only in the 
extreme limit of a very strong electron-phonon coupling, where the polaron 
bandwidth is well below the phonon frequency, can multiphonon structure in $I_{incoh}(E)$
be verified. It has actually been observed in the optical conductivity 
\cite{mul}.

The small Holstein  polaron  is very heavy 
except if the phonon frequency is extremely high \cite{ale2}. In fact, 
oxides are ionic semiconductors where the long-range 
electron-phonon interaction dominates. That leads 
 to a much lighter  mobile small Fr\"ohlich polaron (SFP)  \cite{ale2,alekor}. 
Considering the electron-phonon interaction in a multi-polaron 
system one has to take into account the dynamic properties of the 
dielectric
response function \cite{ale4}. One can naively believe \cite{chaR} that the long-range  
Fr\"ohlich interaction becomes short range (Holstein) 
due to  screening.  This is of course untrue.
In the long-wave limit ($q\rightarrow 0$) the response of polarons at the 
optical phonon frequency is dynamic, because $\omega >> qv$ ($v$ is the 
characteristic group velocity of polarons). Therefore, the singular 
(Fr\"ohlich)
behaviour of $\gamma({\bf q})\sim 1/q$ is unaffected by the 
screening. Polarons 
 are slow enough  and $cannot$  screen the high-frequency 
crystal field oscillations.  Hence, the interaction 
with the high-frequency optical phonons in 
ionic polaron solids remains long-range whatever the carrier
density. It is easy to show \cite{alecha} using Eq.(5) and Eq.(6) that for any 
finite-radius interaction with a $q$-dependent matrix element the 
coherent part of  ARPES spectra takes the same form as Eq.(8), but with a 
$different$ spectral weight ($Z$) and effective mass ($Z'$) renormalisation 
exponents.  Also some ${\bf k}$ dependence of the $incoherent$ 
background appears if   the matrix 
element of the electron-phonon interaction depends on $q$\cite{kay}. Hence, in general, 
\begin{equation}
 I({\bf k},E) \sim |d({\bf k})|^{2}n(E)Z\delta (E+\xi_{\bf 
 k})+I_{incoh}({\bf k},E),
 \end{equation} 
 with the same $Z=exp(-E_{p}/\omega)$ as in the case of the Holstein 
 polaron but with the SFP  bandwidth much less reduced, $\xi_{\bf 
 k}=Z'E({\bf k})-\mu$, where $Z'=exp(-\gamma E_{p}/\omega)$. In general 
 one finds
 \begin{equation}
 \gamma= \sum_{\bf q} |\gamma({\bf q})|^{2} [1-cos ({\bf q \cdot a})]/
\sum_{\bf q} |\gamma({\bf q})|^{2},
\end{equation}
and $E_{p}=(1/2)\sum_{\bf q}|\gamma({\bf q})|^{2}\omega_{\bf q}$.
 The numerical coefficient $\gamma$ might be as small as 
 $0.4$\cite{alekor}
and 
even smaller in the 
cuprates with the nearest neighbour oxygen-oxygen distance less than the 
lattice constant, 
$\gamma \simeq 0.2$ \cite{ale2}, which is precisely 
confirmed by  exact QMC  calculation of the small polaron mass 
\cite{alekor}. On the one hand this important result tells us that small polarons 
as well as   intersite bipolarons are perfectly mobile and can account for 
the high-T$_{c}$ values in cuprates \cite{alecom}. On the other the
 coherent spectral weight remains strongly suppressed 
in polaronic conductors, Eq.(9), because $Z$ might be less than $Z'$
 by one or even a few orders of magnitude. These unusual SFP 
 spectral features  provide an explanation for the apparent discrepancy 
 between a very small Drude weight and a relatively moderate  
  mass enhancement, $m^{*}\sim 3m_{e}е-10 m_{e}е$  (depending on doping) 
 of carriers in manganites \cite{cmr,des} 
 and cuprates. They also explain why the  evHs observed in ARPES 
 \cite{gof,mat,ino} can be hardly seen in angle averaged 
 photoemission. Indeed, the integrated spectral weight of the 
 incoherent background is proportional to $(1-Z)$, Eq.(7). It turns out to be much 
 larger than 
 the coherent contribution, proportional to $Z<<1$. Finally, the ${\bf k}$ 
 dependent $incoherent$ background might obscure the experimental 
 determination of  the Fermi-level crossing, leaving the 
 `leading edge'  determination of it \cite{rand} 
 rather unreliable. 
 
 In the following we concentrate on the angular, spectral and 
 polarisation dependence of the first  coherent term, Eq.(9). The 
 present experimental resolution \cite{rev} allows probing of the 
 intrinsic damping of the coherent polaron tunnelling. This damping 
 appears due 
  to the random field and low-frequency lattice and spin fluctuations 
  described by the polaron self-energy $\Sigma_{p}е ({\bf k},E)$, so that
the coherent part of the spectral function is given by
 \begin{equation}
 A_{p}е({\bf k},E) =-(1/\pi) {\Im \Sigma_{p}е ({\bf k},E)\over{[E+\Re 
 \Sigma_{p}е 
 ({\bf k},E)-\xi_{\bf k})]^{2}+[\Im \Sigma_{p}е ({\bf k},E)]^{2}}}.
 \end{equation}
 Hence, the theory of the narrow ARPES peak reduces to the determination 
 of the  self-energy of the coherent small hole polaron scattered by 
 impurities, low-frequency deformation  and   
 spin fluctuations. 
 
        \centerline {\bf 3. Self-energy of 1D hole in the} 
        \centerline{\bf non-crossing approximation}

Due to energy conservation small polarons exist in the Bloch states 
at temperatures well below the optical phonon frequency $T<<\omega/2$ 
no matter how strong
  their interaction with phonons is
 \cite{lan,hol,app,fir,alemot}. This textbook 
 result, known for a long time, has been  questioned by some 
 authors \cite{ran}. It has been shown \cite{fir2}  that the confusion is 
 due to a profound misunderstanding of the strong-coupling expansion 
  by those authors. The finite polaron self-energy appears only due to the
  (quasi)elastic scattering. First we apply the simplest non-crossing 
  (ladder) approximation to derive the analytical results, Fig.2. Within this 
  approximation the self-energy is ${\bf k}$ -independent for a short-range 
  scattering potential like the deformation or a screened impurity 
  potential, so that
  \begin{equation}
  \Sigma_{p}е(E)  \sim \sum_{\bf k}G_{p}^{R}({\bf k},E),
  \end{equation}
 with $G_{p}^{R}({\bf k},E)=[E-\xi_{\bf k}-\Sigma_{p}е(E)]^{-1}$
 
  The 
  oxygen polaron
  spectrum is parametrised in the tight-binding model 
   as \cite{ale2}
  \begin{equation} 
 \xi^{x,y}_{\bf k}=2[tcos( k_{x,y}a)-t'cos( 
k_{y,x}a)+t_{c}cos(k_{z}d)]-\mu,
\end{equation}
If the oxygen hopping integrals in Eq.(13), reduced by the narrowing effect,  are 
 positive,  the minima of the polaron 
 bands are found at the Brillouin zone boundary in  X $(\pi,0)$ 
and Y $(0,\pi)$.  
  The wave vectors corresponding to the energy minima belong
to the stars with two prongs. Their group has only 1D representations. This
means that the spectrum is degenerate with respect to the number of prongs
of the star. The spectrum Eq.(13) belongs to the star with two prongs, and,
hence it is a two-fold degenerate \cite{alecom2}. The doublet is degenerate if the hole 
resides on the apical oxygen \cite{ref2}. In general, the degeneracy can 
be removed due to the chains in the buffer layers of Y123 and Y124, so 
that the $y$-polaronic band corresponding to the tunneling along the chains
 might be the lowest one.

As  mentioned above the oxygen hole is (quasi) one-dimensional 
  due to a large difference between the oxygen hopping integrals for the 
  orbitals elongated parallel to and perpendicular to the oxygen-oxygen hopping 
  $t',t_{c} << t$.  This allows us to apply a one-dimensional 
  approximation, reducing Eq.(13) to two 1D parabolic bands near the $X$ and $Y$ 
  points,  $\xi^{x,y}_{\bf k}=k^2/2m^{*}-\mu$  with $m^{*}е=1/2ta^2$ and $k$ 
  taking relative to $(\pi,0)$ and $(0,\pi)$, respectively. Then, the 
  equation for the self-energy in the non-crossing approximation, Eq.(12) takes the following form
  \begin{equation}
  \Sigma_{p}е(\epsilon)= -2^{-3/2} [\Sigma_{p}е(\epsilon)-\epsilon]^{-1/2},
  \end{equation}
 for each doublet component. Here we introduce 
 a dimensionless energy (and self-energy), 
  $\epsilon\equiv (E+\mu)/\epsilon_{0}$ using   
  $\epsilon_{0}=(D^{2}m^{*})^{1/3}$ as  the energy unit.
The constant $D$ is the second moment of the Gaussian white noise potential, 
comprising  thermal and random fluctuations as 
$D=2(V_{0}^{2}T/M+n_{im}v^{2})$, where $V_{0}$ is the amplitude of 
the deformation  potential, $M$ is the elastic modulus, $n_{im}$ is the 
impurity density, and $v$ is the coefficient of the $\delta$- 
function impurity potential. The solution is
\begin{eqnarray}
\Sigma_{p}е(\epsilon)&=&{\epsilon\over{3}} 
-\left 
({1+i3^{1/2}\over{2}}\right)\left[{1\over{16}}+{\epsilon^{3}\over{27}}+\left ({1\over{256}}+
{\epsilon^{3}\over{216}}\right )^{1/2}\right ]^{1/3}\cr
&-&\left ({1-i3^{1/2}\over{2}}\right )\left 
[{1\over{16}}+{\epsilon^{3}\over{27}}-\left ({1\over{256}}+
{\epsilon^{3}\over{216}}\right )^{1/2}\right ]^{1/3}.
\end{eqnarray}

While the energy resolution in the present ARPES studies is almost 
perfect \cite{rev}, the momentum resolution remains rather poor, $\delta 
> 0.1 \pi/a$. Hence we have to integrate the spectral function, Eq.(11), 
with a Gaussian momentum resolution  to obtain the experimental 
photocurrent,
\begin{equation}
I({\bf k},E) \sim Z \int_{-\infty}^{\infty} dk' A_{p}е(k',-E) exp 
[-{(k-k')^{2}\over{\delta^{2}}}].
\end{equation}
The integral is expressed in terms of $\Sigma_{p}е(\epsilon)$, Eq.(15)
and the tabulated  Error function $w(z)$ 
 as
\begin{equation}
I({\bf k},E) \sim -{2Z\over{\delta}} \Im \left 
(\Sigma_{p}е(\epsilon)[w(z_{1})+w(z_{2})]\right),
\end{equation}
where $z_{1,2}=[\pm k-i/2\Sigma_{p}е(\epsilon)]/\delta$, $w(z)=e^{-z^{2}} er\!fc(-iz)$ and 
$\epsilon=(-E+\mu)/\epsilon_{0}$. This photocurrent is plotted as
dashed lines in 
Fig.3 for two momenta, $k=0.04 \pi/a$ 
(almost Y or X points of the Brillouin zone) and $k=0.3 
\pi/a$. The chemical potential is placed in the charge transfer gap below 
the bottom of the hole band, $\mu=-20 meV$, the momentum resolution is 
taken as $\delta=0.28 \pi/a$ and the damping $\epsilon_{0}=19 meV$. 

The imaginary part of the self-energy, Eq.(15) disappears below 
$\epsilon=-3/ 2^{5/3}\simeq -0.9449$. Hence this approximation gives a well-defined gap 
rather than a pseudogap. Actually, the non-crossing approximation fails to describe the 
localised states inside the gap (i.e. the Lifshitz tail of the density of 
states). One has to go beyond the simple ladder, Fig.2, to describe the 
single-electron tunnelling 
inside the gap \cite{aletun} 
and the ARPES spectra at very small binding energy.  

 \centerline {\bf 4. Exact spectral function of 1D hole}

The 
exact  spectral function  for a one dimensional particle in 
a                  random Gaussian white noise potential 
was derived by Halperin \cite{hal} and the density of states
 by Frisch and 
Lloyd \cite{fri}. Halperin derived two pairs of differential equations
from whose solutions the spectral function of a 
`Schrodinger' particle (i.e. in the effective mass approximation) and of a 
`discrete' particle (tight-binding approximation) may be calculated. The QMC polaronic 
bandwidth is about $100$ meV or larger \cite{alekor}, which allows us to
 apply  the `Schrodinger' particle spectral function, given by \cite{hal}
\begin{equation}
A_{p}е(k,\epsilon)= 4\int_{-\infty}^{\infty}p_{0}(-z) \Re p_{1}(z)dz,
\end{equation}
where $p_{0,1}(z)$ obey the two differential equations
\begin{equation}
\left[{d^2\over{dz^2}}+{d\over{dz}}(z^2+2\epsilon)\right]p_{0}=0,
\end{equation}
and
\begin{equation}
\left[{d^2\over{dz^2}}+{d\over{dz}}(z^2+2\epsilon)-z-ik\right]p_{1}+p_{0}=0,
\end{equation}
with boundary conditions
\begin{equation}
\lim_{z\rightarrow \infty}z^{2-n}p_{n}(z)=\lim_{z\rightarrow 
-\infty}z^{2-n}p_{n}(z)
\end{equation}
where $k$ is measured in units of $k_{0}=(D^{1/2}m^{*})^{2/3}$. 
The first equation may be integrated to give
\begin{equation}
p_{0}(z)={exp(-z^{3}/3 -2z\epsilon) \int_{-\infty}^{z} exp(u^{3}/3+2u\epsilon) 
du\over{\pi^{1/2}\int_{0}^{\infty} u^{-1/2} exp(-u^{3}/12 -2u\epsilon) 
du}}.
\end{equation}
The equation for $p_{1}(z)$ has no known analytic solution, and hence
must be solved numerically. There is however an asymptotic expression
for $A_{p}е(k,\epsilon)$ in the tail where $|\epsilon|\gg 1$:
\begin{equation}
A_{p}е(k,\epsilon)\sim2\pi(-2\epsilon)^{1\over{2}}exp[-{4\over{3}}(-2\epsilon)^{3\over{2}}]cosh^{2}\left[{\pi
  k\over{(-8\epsilon)^{1\over{2}}}}\right].
\end{equation}
In practice, for computational efficiency we use the exact spectral
 density for $-1.4\leq\epsilon<1$, and outside this range we use the
 asymptotic result for $\epsilon<-1.4$, Eq.(23) and the non-crossing
 approximation for $\epsilon\geq1$, where they are almost
 indistinguishable from the exact result on the scale of the diagrams
 plotted here.

 The result for $A_{p}е(k,-E)$ integrated with the Gaussian 
momentum resolution is shown in Fig.3 for two values of 
the momentum (solid lines). Quite differently from the non-crossing approximation the 
exact spectral function (averaged with the momentum resolution 
function)  has the Lifshitz tail due to the states localised by 
disorder within the normal state gap.  However, besides this tail the 
non-crossing approximation gives very good agreement, and for binding
energy greater than about 30meV it is practically exact. 

The cumulative DOS
\begin{equation} 
N_{p}е(\epsilon)=(2\pi)^{-1}\int_{-\infty}^{\epsilon}d\epsilon'
\int_{-\infty}^{\infty}dk A_{p}(k,\epsilon')
\end{equation}
is expressed analytically \cite{fri} in terms of the tabulated Airy 
functions $Ai(x)$ and $Bi(x)$ as
\begin{equation}
N_{p}е(\epsilon)) = \pi^{-2}\left[Ai^{2}(-2\epsilon) +Bi^{2}(-2\epsilon)\right]^{-1}.
\end{equation}
The DOS  $dN(\epsilon)/d\epsilon$
fits well the voltage-current 
tunnelling characteristics of  cuprates \cite{aletun}. 

\centerline {\bf 5. Theory of ARPES in Y124 and Y123}

With the polaronic doublet, Eq.(13) placed above the chemical potential 
we can quantitatively describe high-resolution ARPES in Y123 
\cite{mat} and Y124 \cite{gof}. First we  explain the 
experimentally observed polarisation 
dependence of the intensity near Y and X \cite{mat}. The Bloch 
periodic function $u_{\bf k}({\bf r})$ can be expressed in terms of 
the Wannier orbitals $w({\bf r})$ as
\begin{equation}
u_{\bf k}({\bf r})=N^{-1/2}\sum_{\bf m}e^{i{\bf k} \cdot \bf {(m-r)}} w ({\bf 
r-m}).
\end{equation}
Then the dipole matrix element is given by the derivative of the 
Fourier component of the atomic (Wannier) orbital, $f_{\bf k} \equiv 
v_{0}^{-1/2}\int d{\bf r} w({\bf r}) exp (i{\bf k \cdot r})$ее as
\begin{equation}
{\bf d}({\bf k}) =i({\bf e}\cdot  \nabla_{\bf k}) f_{ {\bf k}}.
\end{equation}

To estimate $f_{\bf k}$ we approximate the $x,y$ oxygen orbitals 
contributing to the x and y polaronic bands, respectively, 
with $w_{x}е({\bf r})= (1/8a_{0}^{3}\pi)^{1/2}е (x/2a_{0}) exp(-r/2a_{0})$ 
and $w_{y}е({\bf r})=(1/8a_{0}^{3}\pi)^{1/2}(y/2a_{0})е 
exp(-r/2a_{0})$. As a result we obtain for the x orbital,
\begin{eqnarray}
{\partial f_{\bf k}\over\partial k_{x}}&=&(8a_{0}^{3}\pi/v_{0}е)^{1/2}a_{0}\cr
&\times&\left( [(ka_{0})^{2}+1/4]^{-3} -6 (k_{x}a_{0})е
^{2}[(ka_{0})^{2}+1/4]^{-4}\right),
\end{eqnarray}
and
\begin{equation}
{\partial f_{\bf k}\over\partial k_{y,z}}=-6(8a_{0}^{3}е\pi /v_{0})^{1/2}a_{0}^{3}
k_{x}k_{y,z}е[(ka_{0})^{2}+1/4]^{-4},
\end{equation}
Here ${\bf k}$ is the photoelectron momentum and $a_{0}$ is the size 
of the Wannier function. For the case of y-orbital one should interchange x and y. Near the X and Y 
points of the Brillouin zone, $|k_{y,x}|<<k$, respectively. Then it 
follows from Eq.(28) and Eq.(29)  that
 the ARPES peak should be seen at X and almost disappear at Y
if the photons are polarised along the x-direction, i.e. ${\bf e}\parallel
\bf a$. If the polarisation is along the y-direction 
 (${\bf e}\parallel {\bf b}$) the peak appears at Y  and almost 
 disappears at X. Precisely this behaviour is observed in 
 ARPES spectra obtained using polarised photons\cite{mat}, Fig.4.
  We also notice a very strong dependence of the dipole matrix element, 
 Eq.(28) 
 on the photon energy, $d\sim \nu^{-3}$ at large $\nu$. Hence, it is not surprising 
 if the ARPES peak disappears at large $\nu$ as has been recently 
 observed \cite{des2}.
 
The exact 1D polaron spectral function, Eq.(18),  integrated with the 
 experimental momentum resolution,  provides a quantitative fit 
 to the ARPES spectra in Y124   along the $Y-\Gamma$ direction, as shown 
 in Fig.5. The angular dispersion is described with the polaron mass 
 $m^{*}= 9.9 m_{e}$ in  agreement with the Monte-Carlo calculations of 
 the SFP mass \cite{alekor}. The spectral shape is 
 reproduced well with $\epsilon_{0}= 19 meV$, Fig.6, in close
 agreement with 
 the value of this parameter found in tunnelling experiments 
 \cite{aletun}. 
 That yields an estimate of the polaron scattering rate, which 
 appears to be smaller than the polaron bandwidth (about $100 meV$ 
 or larger), in agreement with the 
 notion \cite{ale2} that many high-T$_{c}$ cuprates are in the clean 
 limit.  There is also quantitative agreement in the perpendicular 
 direction  $Y-S$, Fig.7, in a restricted region of small $k_{x}$,
  where almost no dispersion  is observed 
 around Y. Slight dispersion in the $Y-S$ direction towards the 
 chemical potential might be due to a negative $t'$ in Eq.(13). 
 
 However, there is a significant loss of the energy-integrated 
 intensity along both directions, Fig. 8, which the theoretical 
 spectral function alone cannot account for.
 The energy-integrated ARPES spectra obey the sum rule,
 \begin{equation}
 \int_{-\infty}^{\infty} dE  I({\bf k},E) \sim |d({\bf k})|^{2}n_{\bf k},
 \end{equation}
 where $n_{\bf k}=\langle h_{\bf k} h^{\dagger}_{\bf k}\rangle$.
 If the dipole matrix element is almost k-independent and 
 the chemical potential is pinned well inside the charge-transfer gap,
  so that $n_{\bf k}=1$  
  this integral would be ${\bf k}$ independent as well. This is not 
  the case for Y124, no matter what  the scanning direction is, Fig.8. 
  Therefore, we have to conclude that either the dipole matrix element 
  is ${\bf k}$ dependent or (and) the oxygen band is strongly 
  correlated (in the Mott-Hubbard sense).  As we have mentioned above, 
  the incoherent background of SFP is angle dependent as well, which might 
   contribute to the intensity loss. 
  
  The rapid loss of the integrated intensity in the 
  $Y-S$ direction was interpreted by Randeria and Campuzano 
  \cite{rand} as a Fermi-surface crossing. While a Fermi-surface 
  crossing is not  incompatible with our scenario  (see Fig.1 inset), 
  we do not believe that it has really been observed in Y124. First of all 
  these authors suppressed a few experimental curves in the 
  $Y-\Gamma$ direction, Fig.5, which prevented them from observing the 
  intensity loss in this `dielectric' direction, where there is
  obviously no Fermi-surface crossing.  This loss of  intensity along
   $Y-\Gamma$ tells us that the intensity loss might be  due to the matrix element 
  rather than to the Fermi-surface 
  crossing in both directions. This is confirmed by our observation of 
  a similar rapid loss of the intensity in a $dielectric$, FeSi 
  \cite{par}, Fig.9 with no Fermi-surface at all. The peaks in the $Y-S$ direction are all 15 meV or more below the
  Fermi level - at a temperature of 1 meV, if the loss of spectral
  weight were due to a Fermi-surface crossing one would expect the
  peaks to approach much closer to the Fermi level. Also the 
  experimental spectral 
  shape of the intensity at  ${\bf k}={\bf 
  k}_{F}$ is incompatible with any theoretical scenario, including different marginal 
  Fermi-liquid models, Fig.10. The spectral function on the 
  Fermi surface should be close to a simple Lorentzian,
 \begin{equation}
  A_{p}е({\bf k}_{{F}}е,E)\sim {|E|^{\beta}е\over{E^{2}
  +constant \times E^{2\beta}}},
 \end{equation}
because the imaginary part of the self-energy behaves as 
$|E|^{\beta}$ with $0\leq \beta \leq 2$. On the contrary, the 
experimental intensity shows a pronounced minimum at the alleged 
Fermi-surface, Fig.10.  

If there is indeed no Fermi-surface crossing in many cuprates, as we 
argue,  why 
then does the `maximum locus' determination  point to a large Fermi 
surface in cuprates, which is drastically incompatible with their kinetic and 
thermodynamic properties? We propose that it appears due to the fact 
that oxygen semiconducting band has its minima at large $k$ inside or 
even on the boundary of the Brillouin zone. That is why 
ARPES  show intense peaks near large $\bf k$ imitating a large 
Fermi-surface.

  \centerline {\bf 6. Summary and conclusions} 
  
  In summary, we have proposed a theory of ARPES in cuprates based 
  on the LDA+U band structure and the bipolaron theory compatible with 
  the normal state kinetic and thermodynamic properties of these 
  materials. The theory explains the narrow flat bands observed in Y123 
  and Y124, including their polarisation, spectral and angular 
  dependence, as well as a featureless (but dispersive) background. 
  The ARPES peak originates from  the hole excitations of the 
  polaronic oxygen band of the buffer layers, in  agreement with 
  the experimental results and electronic structure of Schabel $et$ 
  $al$ \cite{mat}. Differently from these authors we suggest that this 
 band is intrinsic for cuprates and takes part in the bipolaron 
 formation and superconductivity, which is nicely confirmed by  a few independent 
 studies \cite{ref2}.   The normal state gap 
  is  half of the bipolaron binding energy. The angular dependence 
  of the peak and of the gap is due to  the polaron band 
  dispersion, which agrees well with the QMC results for the small 
  Fr\"ohlich polaron. 
  
  The spectral shape of the peak is affected by the soft 
  lattice, spin and random fluctuations. The characteristic 
  scattering rate agrees well with that found in the tunnelling 
  experiments\cite{aletun}. This scattering rate is temperature 
  dependent not only due to the thermal lattice fluctuations,  but also 
  because of the anomalous screening below T$_{c}$ in the charged 
  Bose-liquid\cite{alemot}. The Bose-Einstein condensate screens 
  effectively the long-range Coulomb potential of impurities. As a 
  result one can expect a drastic change of the damping $\epsilon_{0}$ 
  when T$_{c}$ is passed. That might help to understand the near
  disappearance of the narrow peak above T$_{c}$ in some 
  Bi-cuprates. On the other hand in the stehiometric Y124 with 
  (theoretically) no impurities, one can expect about the same results
  from ARPES 
  below and above T$_{c}$, which seems to be  the case \cite{gof}.
  Within the bipolaron theory there is only one $single-particle$ 
  gap, which is half of the bipolaron binding energy both below and 
  above T$_{c}е$.
   
We believe that many cuprates are doped insulators with no Fermi 
surface at all due to the bipolaron formation.
The Fermi-surface crossing, if it were firmly established  in the overdoped 
samples,   would correspond to a small Fermi surface 
of the oxygen band pockets located at finite ${\bf k}$  like in many 
ordinary semiconductors, for example, in Ge and Si.

The authors greatly appreciate enlightening discussions with  
 D.S. Dessau, N.E. Hussey, V.V. Kabanov,  P.E. Kornilovitch, G.J. Kaye, A.I. 
 Lichtenstein,
 G.A. Sawatzky, J.R. Schrieffer,  Z.-X. Shen,
J. Zaanen,  G. Zhao, and R. Zeyher. C.J.D. has been supported in this
 work by a grant from the EPSRC of the UK.

\centerline{{\bf Figure Captures}}

Fig.1. Schematic LDU+U density of states. The chemical potential is 
pinned inside the charge transfer gap (a) due to the bipolaron 
formation in underdoped cuprates. It might enter the oxygen band in 
overdoped cuprates (b) if the polaron band crosses the bipolaron one (inset).

Fig.2.  The non-crossing diagram for the self-energy. The dashed line 
corresponds to the random potential and (or) to the thermal lattice 
and spin fluctuations.

Fig.3. The polaron spectral function, integrated with the momentum 
resolution function for two angles, $k=0.04 \pi/a$ (upper curves), 
and $k=0.30 \pi/a$ with the damping  $\epsilon_{0}=19 meV$, the momentum 
resolution  $\delta=0.28 \pi/a$ and the polaron mass $m^{*}=9.9 m_{e}$. 
The bipolaron binding energy $2|\mu|=40 meV$. The dashed curves are the
spectral density integrated with the momentum resoltion in the non-crossing approximation.

Fig.4. Polarisation dependence of the ARPES peak in Y123\cite{mat}
 near X and Y points.

Fig.5.  Theoretical ARPES spectra (b) compared with experiment (a) in Y124
\cite{gof} for $Y-\Gamma$ direction. Parameters are those of Fig.3.

Fig.6. Theoretical fit (dashed lines) to two experimental ARPES curves 
corresponding to $ k=0.04 \pi/a$ (upper curves), 
and $k=0.30 \pi/a$.

Fig.7. Theoretical ARPES spectra (b) compared with experiment (a) in Y124 
\cite{gof} for $Y-S$ 
direction.

Fig.8. The energy-integrated  ARPES intensity in Y124 in the 
$Y-\Gamma$ (a) and $Y-S$ (b) directions. Momenta are measured relative
to the Y-point of the Brillouin zone.

Fig.9.  (a) The energy-integrated  ARPES intensity in FeSi \cite{par} as a function 
of the analyser angle. The spectra are shown in (b).

Fig.10. The experimental ARPES signal (solid line) on the alleged Fermi-surface does not 
correspond to a  Fermi-liquid spectral 
function (dashed line). We assume particle-hole symmetry to obtain the spectral
function for negative binding energy.

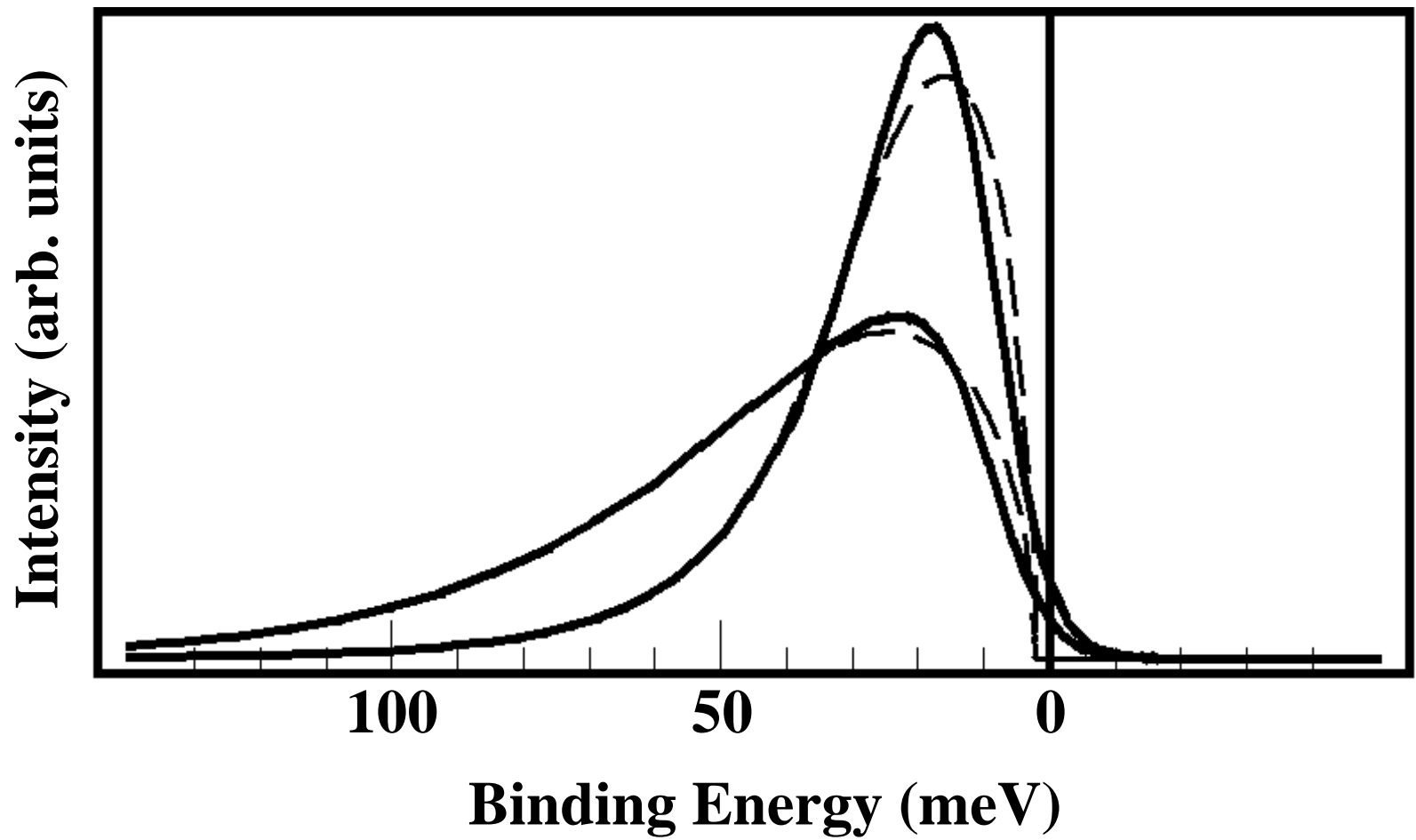
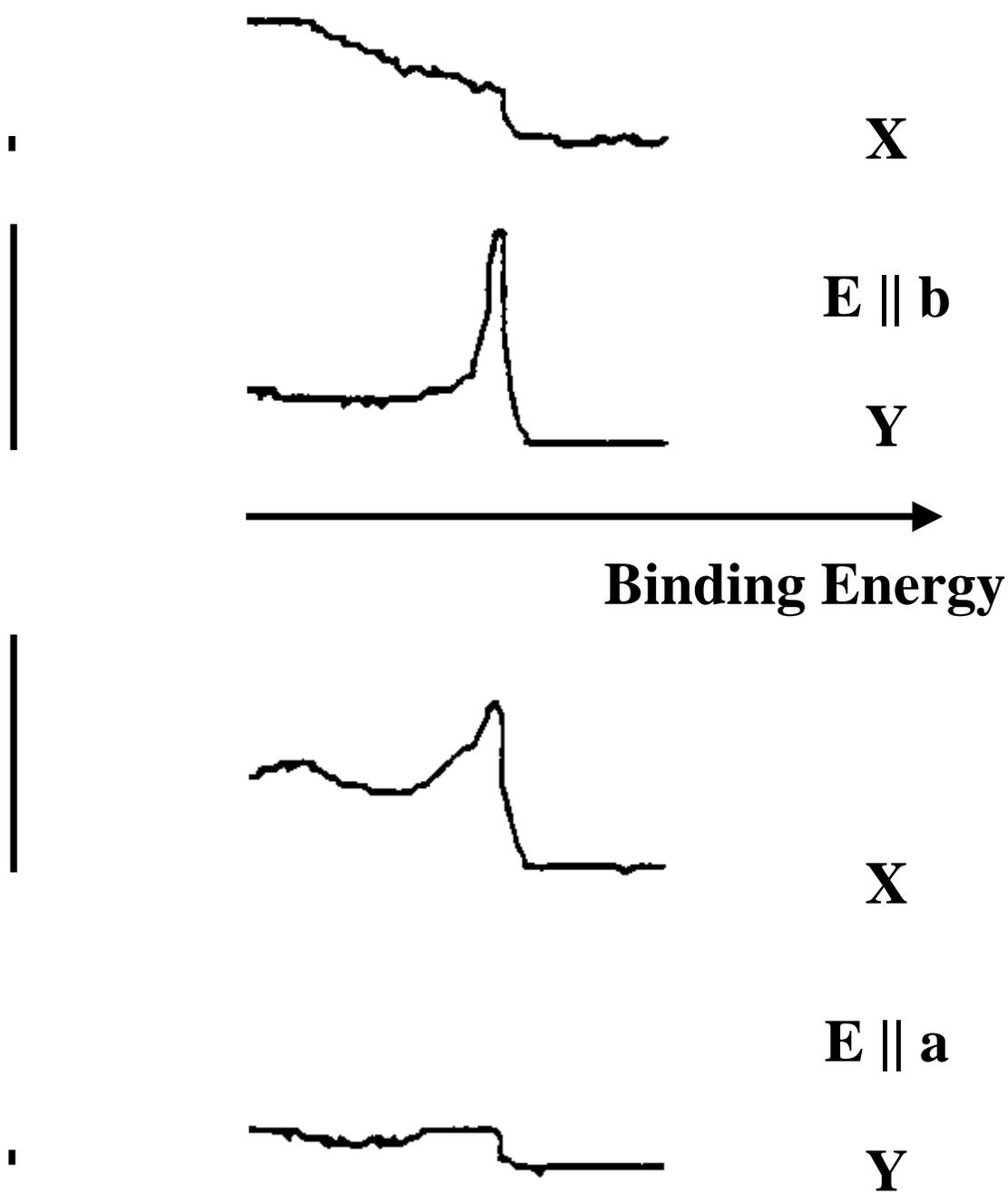
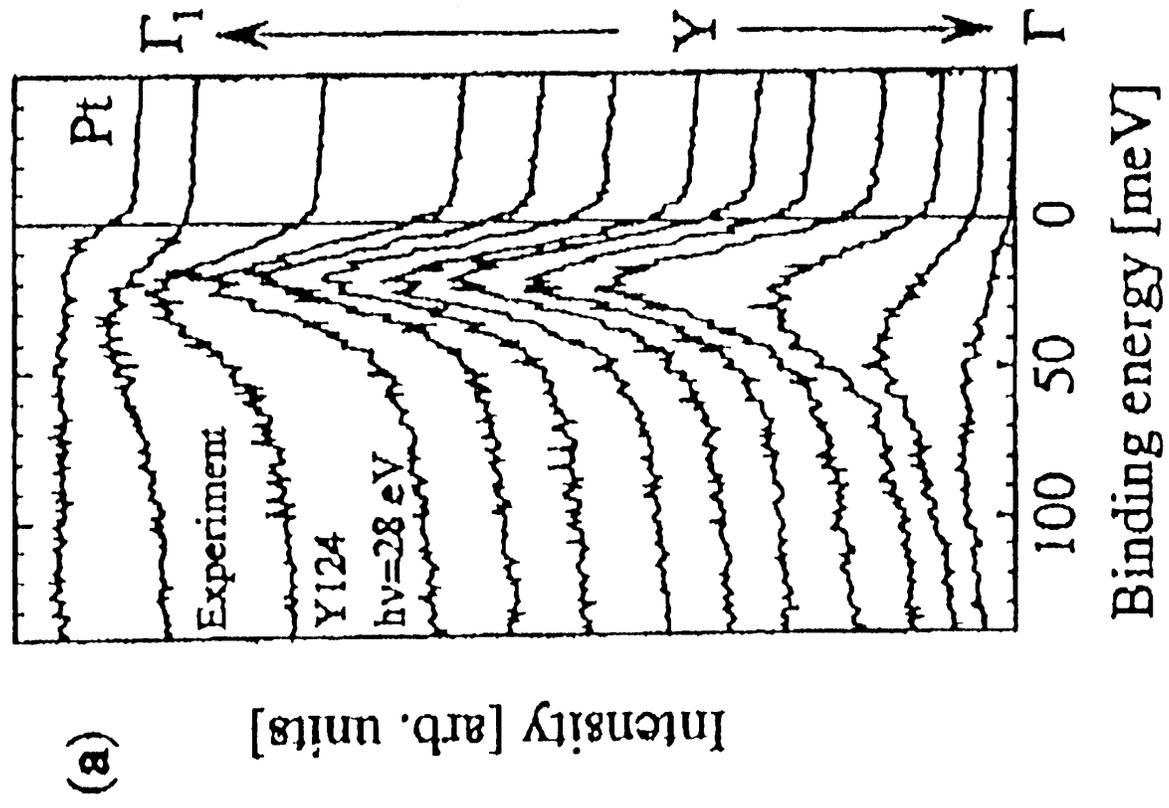
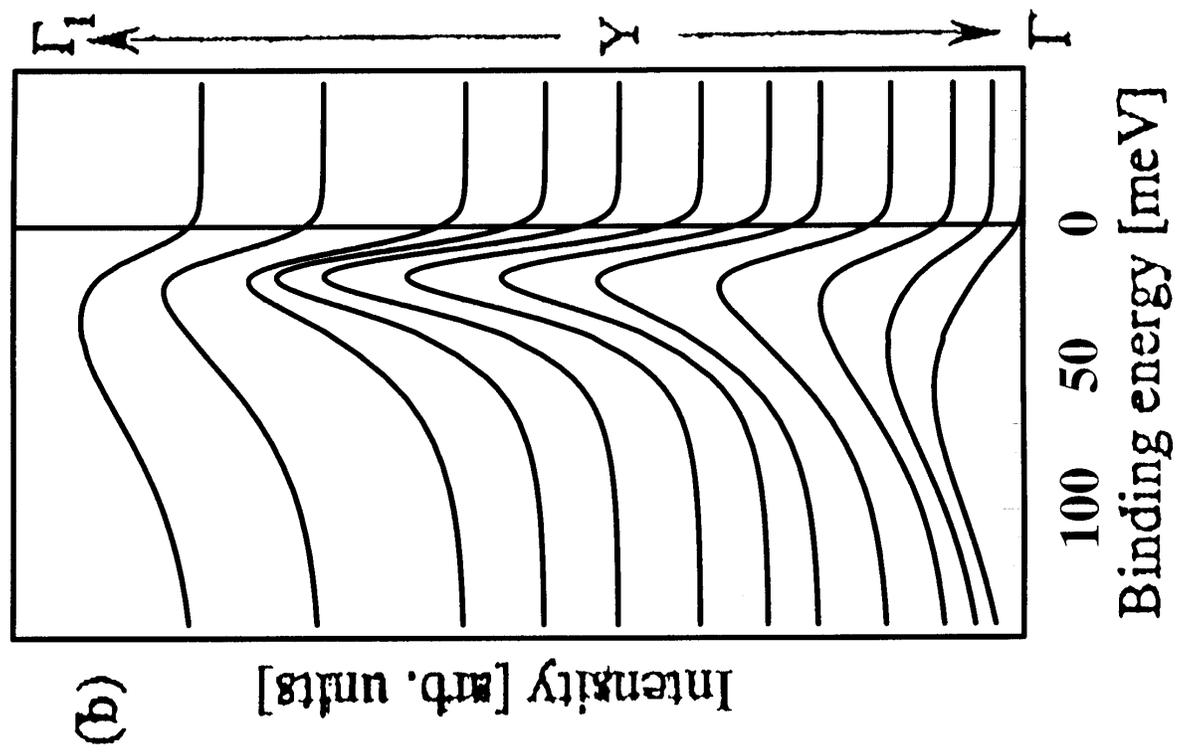
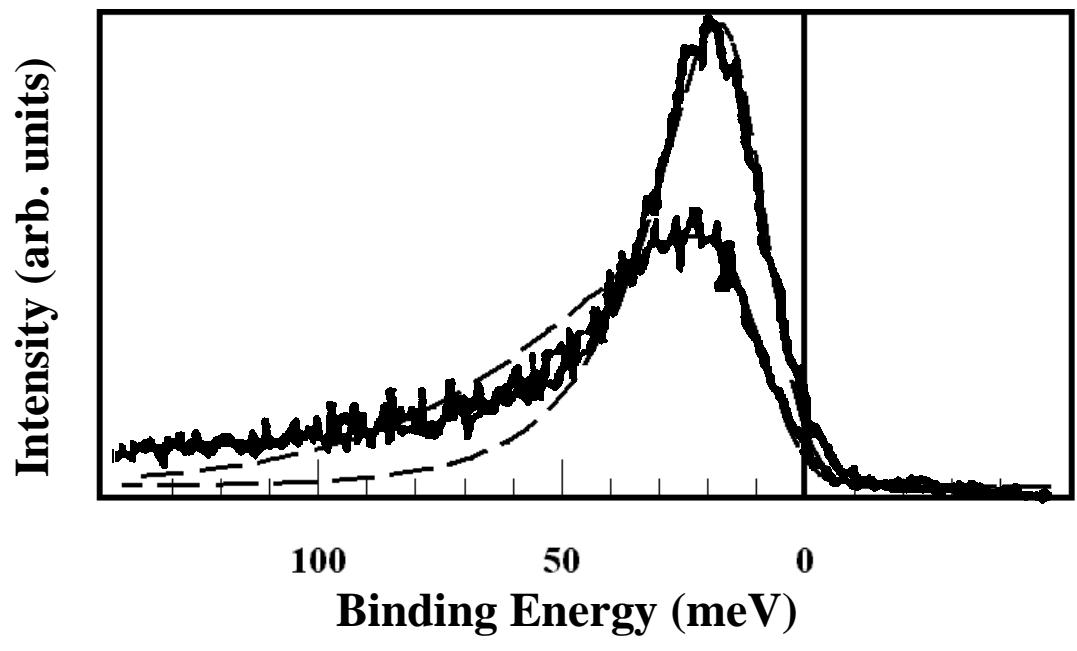
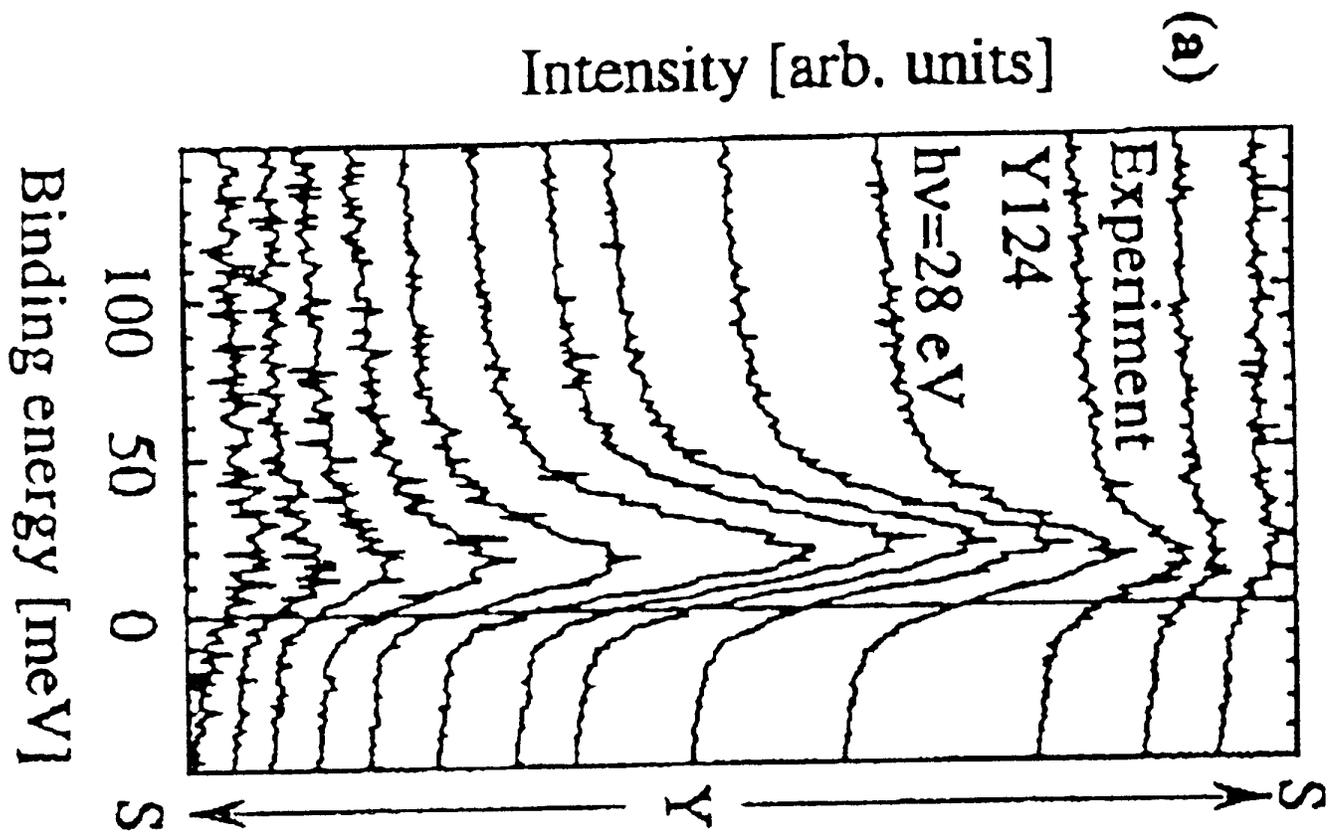
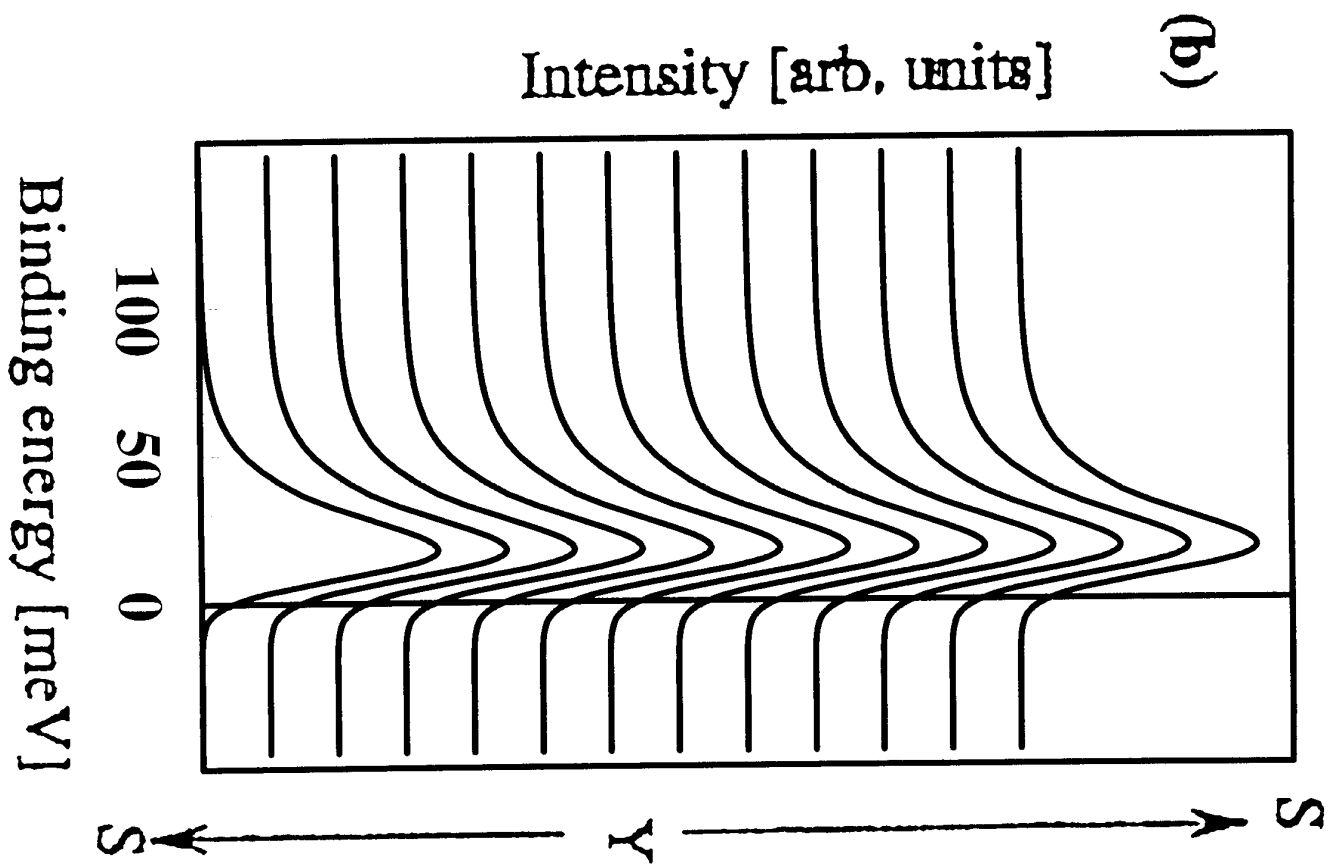
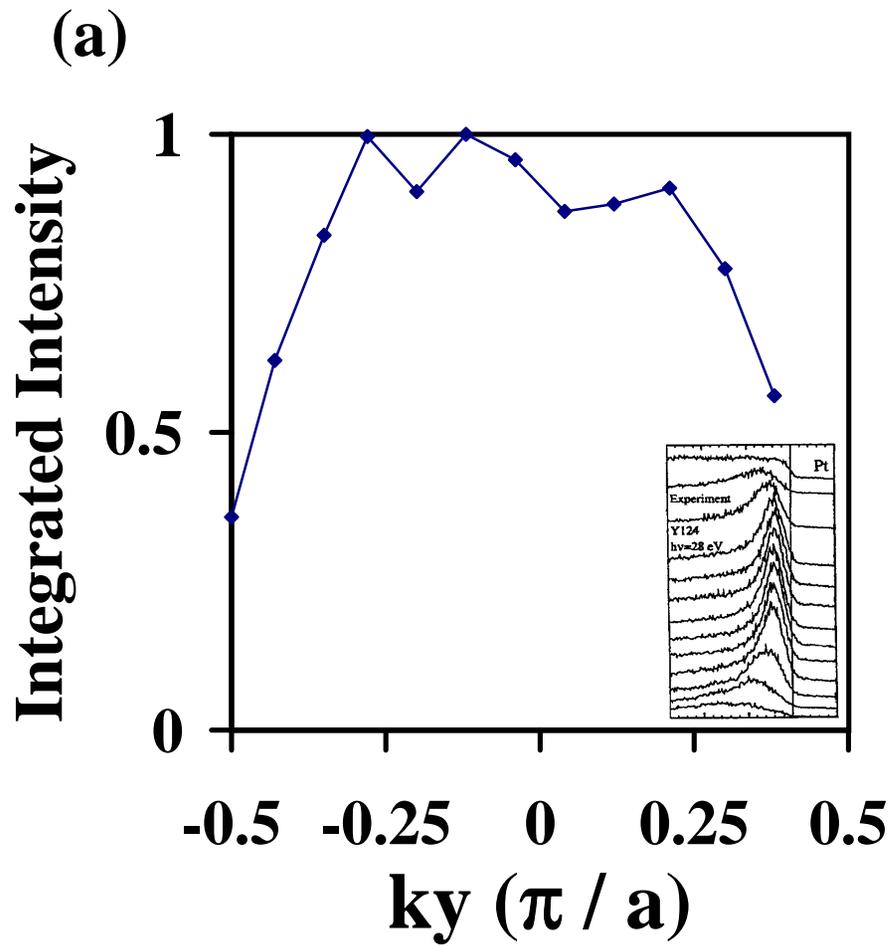
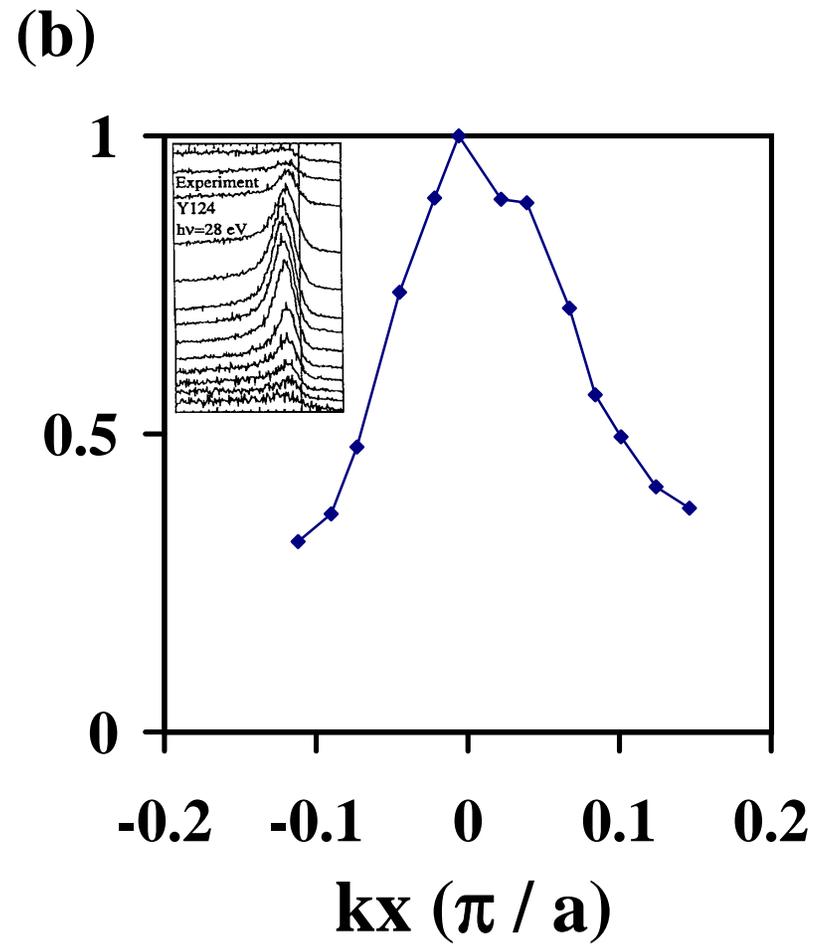
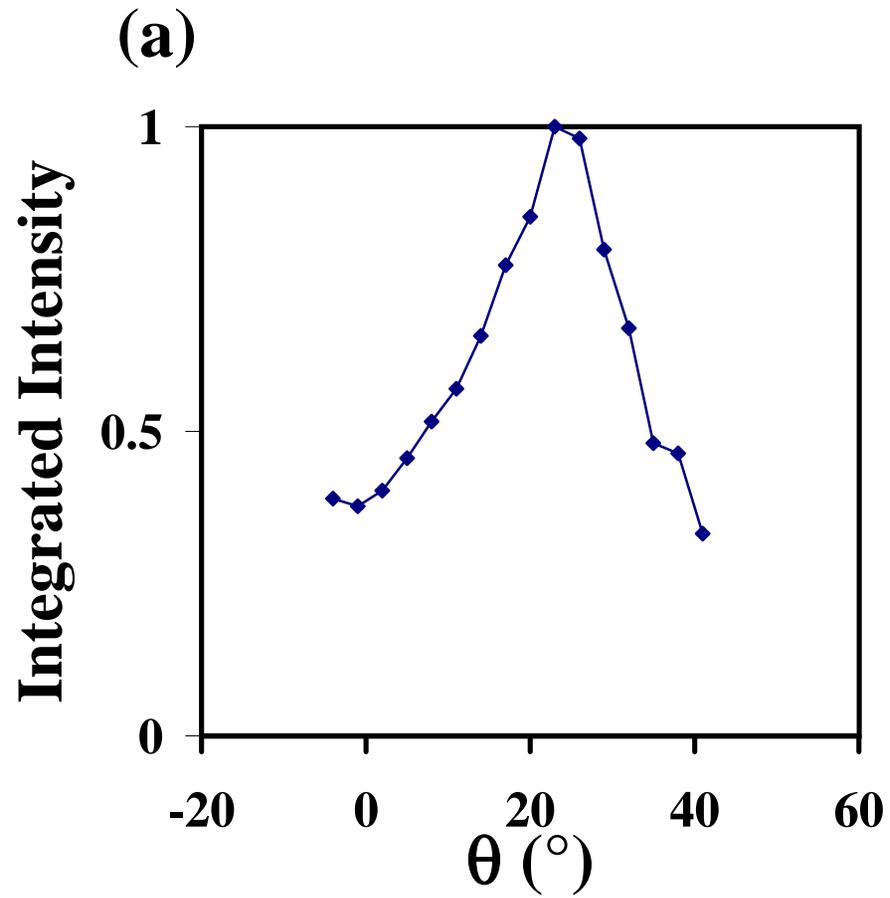
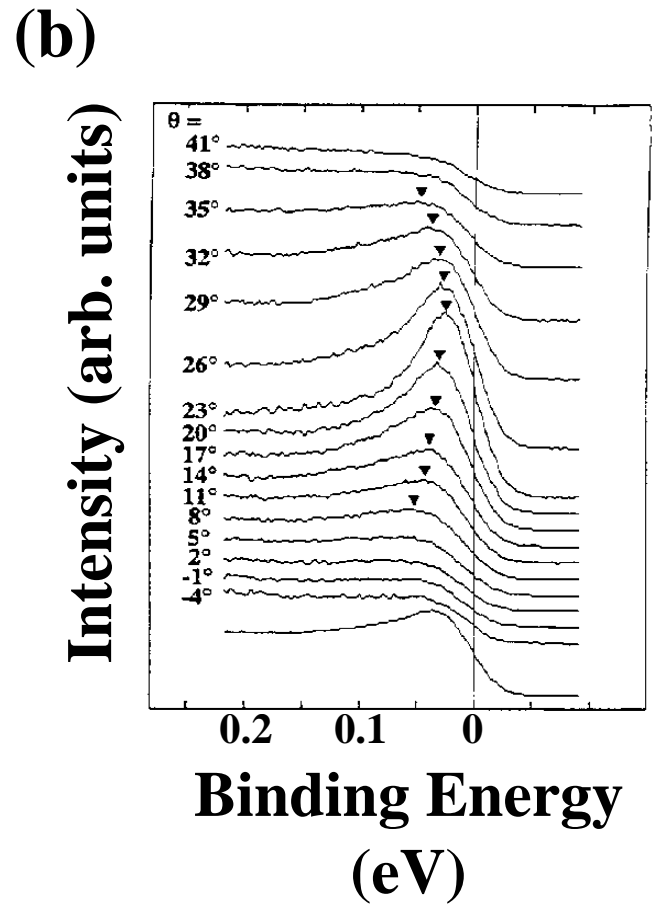
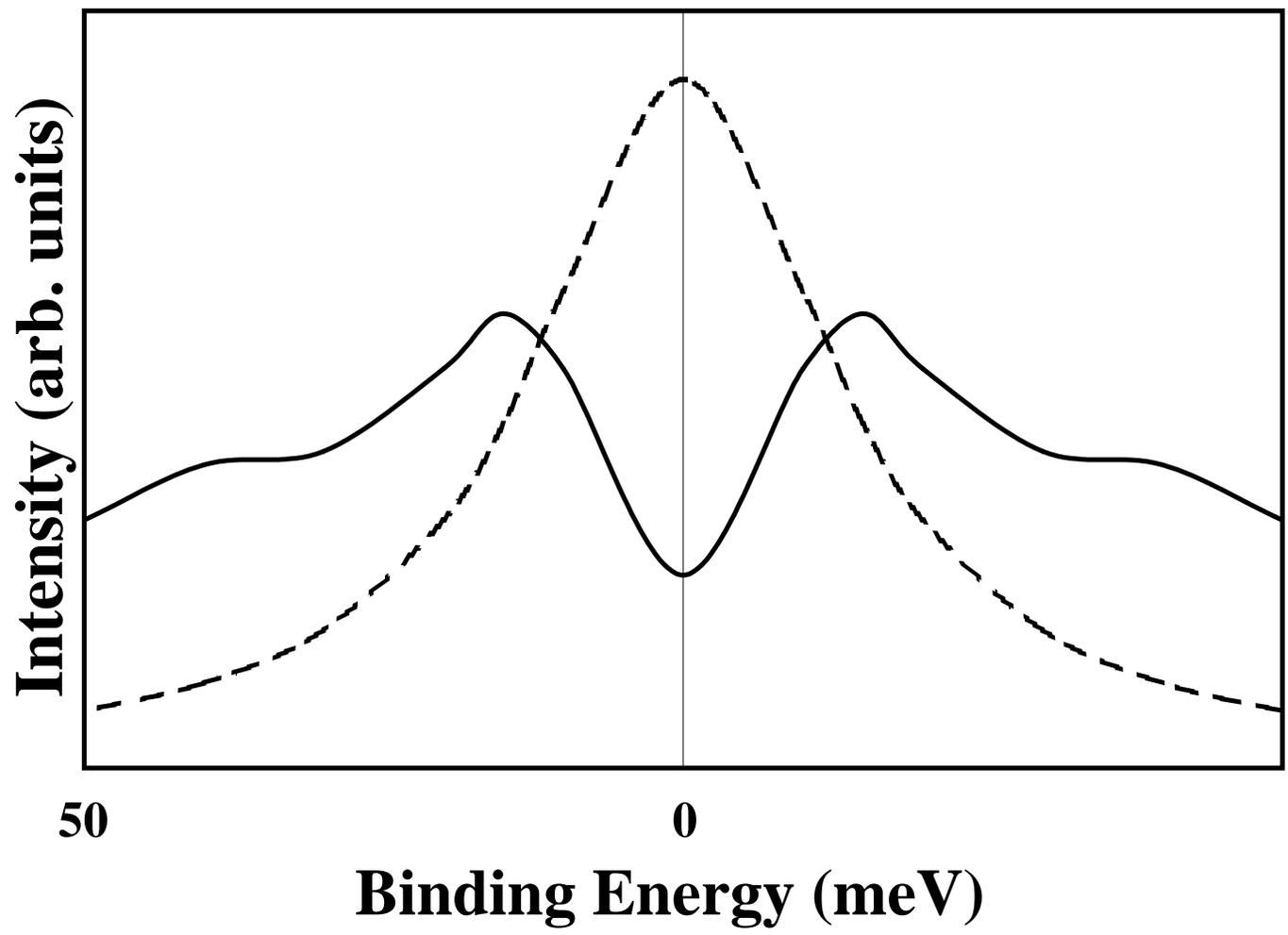
\end{document}